\documentclass[aps,prd,showpacs,nofootinbib,preprintnumbers,twocolumn]{revtex4-1}

\usepackage{bm}
\usepackage{latexsym}
\usepackage{natbib}
\usepackage{url}
\usepackage{dcolumn}
\usepackage{color}
\usepackage{amsfonts,amssymb,amsmath}
\usepackage{graphicx,epsfig}
\usepackage{psfrag}
\usepackage{subfigure}
\usepackage{natbib}

\begin{document}

\title{Cosmological arrow of time in $f(R)$ gravity}


\author{ Bal Krishna Yadav}
\email[]{ balkrishnalko@gmail.com}

\author{Murli Manohar Verma}
\email[]{sunilmmv@yahoo.com}

\affiliation{Department of Physics, University of Lucknow, Lucknow 226 007, India}
\date{\today}

\begin{abstract}
The cosmological arrow of time may be linked to the thermodynamic arrow by second law of thermodynamics. The time asymmetry is also associated with dissipative fluid as Tolman introduced a viscous fluid to generate an arrow of time in cyclic cosmology. An arrow of time in cyclic cosmology has been shown using scalar field.In this work we find out the cosmological arrow of time in f(R) gravity. Here we use the relation between a new scalar field and $f(R)$. The dynamics of this new scalar field may emerge the arrow of time.
\end{abstract}
\maketitle
\section{\label{1}Introduction}
An `arrow' of time is a physical process or phenomenon that has a definite direction in time. The time reverse of such a process does not occur. Eddington thought he had found such an arrow in the increase of entropy in isolated systems. He wrote: The law that entropy always increases - the second law of thermodynamics - holds, I think, the supreme position among the laws of Nature. Since he held the universe to be an isolated system, he thought that its entropy, which he called its `random element', must increase until it reached thermodynamic equilibrium, by which point all life, and even time's arrow itself, must have disappeared. He called this process `the running-down of the universe'. It may be associated with some form of dissipation. Tolman found successive irreversible expansions and contractions in closed universe \cite{a1}. He relates it to the continued increase in the entropy of any selected element of fluid in the model. The cosmological arrow of time has been explained in cyclic Universe without any dissipation in the presence of scalar field\cite{a2}. In this work we find out the arrow of time in $f(R)$ gravity.
\section{\label{2}Conformal transformation}
The action in $f(R)$ corresponds to a nonlinear function of Ricci scalar $R$. In the Jordan frame it is given by:
\begin{eqnarray}\mathcal{A} = \frac{1}{2\kappa^{2}}\int\sqrt{-g}f(R)d^{4}x + \mathcal{A}_{m} \label{a2}\end{eqnarray}
We can rewrite the action (\ref{a2})
\begin{eqnarray}\mathcal{A} = \int\sqrt{-g}\left(\frac{1}{2\kappa^{2}}FR - U \right)d^{4}x + \mathcal{A}_{m},\label{g1}                                \end{eqnarray}
where \begin{eqnarray} U = \frac{FR-f}{2\kappa^2}.\label{g2} \end{eqnarray}
and $\mathcal{A}_{m}$ is the action for relativistic and non-relativistic matter, $\kappa^{2}= 8\pi G$ and $g$ is the determinant of metric tensor $g_{\mu\nu}$.

We consider the field equations in the background of spatially flat Friedmann-Lemaitre-Robertson-Walker (FLRW) spacetime with a metric
\begin{eqnarray}ds^{2}= -dt^{2} + a^2(t)[dr^2 + r^2 (d\theta^{2} + \sin^{2}\theta d\phi^{2})]\label{a0} \end{eqnarray}
where $a(t)$ is time dependent scale factor.

It is possible to derive an action in the Einstein frame under the conformal transformation
\begin{eqnarray} \tilde{g}_{\mu\nu} = \Omega^2g_{\mu\nu},\label{g3}\end{eqnarray}
where $\Omega^2$ is the conformal factor and a tilde represents quantities in the Einstein frame. Relation between Ricci scalars in two frames is
\begin{eqnarray} R = \Omega^2(\tilde{R} + 6\tilde{\Box}\omega - 6\tilde{g}^{\mu\nu}\partial_{\mu}\omega\partial_{\nu}\omega),   \label{g4}\end{eqnarray}
where\begin{eqnarray} \omega \equiv \ln\Omega,     \partial_{\mu}\omega\equiv\frac{\partial\omega}{\partial\tilde{x}^{\mu}},    \tilde{\Box}\omega \equiv\frac{1}{\sqrt{-\tilde{g}}}\partial_{\mu}(\sqrt{-\tilde{g}}\tilde{g}^{\mu\nu}\partial_{\nu}\omega).\label{g5}\end{eqnarray}
Now the action (\ref{g1}) is transformed as \begin{eqnarray} \mathcal{A} = \int d^{4}x \sqrt{-\tilde{g}}\left[\frac{1}{2\kappa^{2}}F\Omega^{-2}(\tilde{R} + 6\tilde{\Box}\omega - 6\tilde{g}^{\mu\nu}\partial_{\mu}\omega\partial_{\nu}\omega) - \Omega^{-4}U \right] \nonumber\\ + \mathcal{A}_{m}. \label{g6}\end{eqnarray}
It is possible to write the linear action in $\tilde{R}$ for the choice
\begin{eqnarray} \Omega^{2} = F. \label{g7}\end{eqnarray}
Let us consider a new scalar field $\phi$ defined by
\begin{eqnarray} \kappa\phi \equiv \sqrt{\frac{3}{2}}\ln F. \label{g8}\end{eqnarray}
Now using these relations we get the action in Einstein frame is
\begin{eqnarray}\mathcal{A} = \int d^{4}x \sqrt{-\tilde{g}}\left[\frac{1}{2\kappa^{2}}\tilde{R} -  \frac{1}{2}\tilde{g}^{\mu\nu}\partial_{\mu}\phi\partial_{\nu}\phi - V(\phi) \right] + \mathcal{A}_{m}.  \label{g9}\end{eqnarray}
where \begin{eqnarray} V(\phi) = \frac{U}{F^{2}} = \frac{FR -f}{2\kappa^2F^2}.   \label{g10}\end{eqnarray}
\section{\label{3}Arrow of time}
Let us consider the dynamics in the Einstein frame in the absence of matter fluids. We have the field equation-
\begin{eqnarray} \frac{d^{2}\phi}{d\tilde{t}^{2}} + 3\tilde{H}\frac{d\phi}{d\tilde{t}} + V_{,\phi} = 0. \label{g11}\end{eqnarray}
The energy density and pressure of a homogeneous scalar field are, respectively,
\begin{eqnarray} \rho = \frac{1}{2}\dot{\phi}^{2} + V(\phi), p = \frac{1}{2}\dot{\phi}^{2} - V(\phi), \label{g12}\end{eqnarray}
and the scalar field equation of motion is given by (\ref{g11}).

Tolman described a cyclic universe with progressively larger cycles, assuming the presence of a viscous fluid with pressure
\begin{eqnarray} p = p_{0} - 3\zeta H, \label{g13}\end{eqnarray}
where $p_{0}$ is the equilibrium pressure and $\zeta$ is the coefficient of bulk viscosity. It is clear from equation (\ref{g13}) that $p<p_{0}$ during expansion $(H>0)$ whereas $p>p_{0}$ during contraction. This asymmetry during the expanding and contracting phases results in the growth of both energy and entropy. This increase in entropy makes the amplitude of successive expansion cycles larger leading to a arrow of time.

The term $3\tilde{H}\frac{d\phi}{dt}$ in (\ref{g11})behaves like friction and damps the motion of the scalar field when the universe expands $(H>0)$. In a contracting universe $3\tilde{H}\frac{d\phi}{dt}$ behaves like anti-friction and accelerates the motion of the scalar field. A scalar field with the potential $V = m^{2}\phi^{2}$ gives $p \simeq -\rho$  when $(H>0)$ and $p\simeq\rho$ when $(H<0)$.

In the Friedmann universe with a scalar field the evolution of the model universe is expressed by the trajectories in the three dimensional phase space whose variables are $\phi$, $\dot{\phi} $ and $H$ \cite{a3}. These trajectories can help give some intuition for the cosmology of a particular model defined by the potential $V$. To get more information is to determine all of the critical points. There are finite or infinite critical points. Every trajectory must begin and end at these critical points. In the spatially flat case, the trajectories in three dimensional phase space are confined on a cone.
These results are similar to that of the Tolman.

\section{\label{4}Conclusion}
Here, we get an arrow of time in the absence of matter (relativistic +non-relativistic) in the flat universe in $f(R)$  gravity. This system may be dissipative and this dissipation of scalar field leading irreversibilty of  time. Using this we can explain inflation as well as irreversibility of time in universe. For suitable models present cosmic accelerated expansion can also be explained.

\end{document}